\documentclass[twocolumn,letter]{jpsj2}
\usepackage{graphicx}
\usepackage{amsmath,amssymb}

\title{ 
Non-centrosymmetric Superconductivity and Antiferromagnetic Order: \\
Microscopic discussion of CePt$_3$Si
} 

\author{Youichi {\sc Yanase}$^{1,2}$\footnote{E-mail:
yanase@itp.phys.ethz.ch} and Manfred {\sc Sigrist}$^{2,3}$}

\inst{1 Department of Physics, University of Tokyo, Tokyo 113-0033, Japan \\
\noindent 2 Theoretische Physik, ETH-Honggerberg, 8093 Zurich, Switzerland \\
\noindent 3 Department of Physics, Kyoto University, Kyoto 606-8502, Japan}

\recdate{Today 2006}

\abst
{ The influence of antiferromagnetic order on the superconductivity 
in the non-centrosymmetric heavy fermion compound CePt$_3$Si and related 
materials is discussed. 
Based on our RPA  analysis for the extended Hubbard model
two phases could be stabilized  by a spin fluctuation induced pairing, 
with either dominantly $p$-wave or $d$-wave symmetry. 
The antiferromagnetic order plays an essential role for the low-energy 
physics, in particular, for the appearance of line nodes in the gap and 
the enhancement of spin susceptibility below $T_{\rm c}$. 
Various properties and possible phase diagrams under pressure are analyzed. 
The present experimental situation suggests that the $p$-wave phase 
is most likely realized in CePt$_3$Si.
 
}

\kword
{
Superconductivity without inversion center; antiferromagnetic superconductor
}

\begin{document}
\sloppy
\maketitle

\newcommand{\eli}{$\acute{{\rm E}}$liashberg }
\renewcommand{\k}{\vec{k}}
\newcommand{\kp}{\vec{k}_{+}}
\newcommand{\kk}{\vec{k'}}
\newcommand{\kkk}{\vec{k''}}
\newcommand{\q}{\vec{q}}
\newcommand{\Q}{\vec{Q}}
\newcommand{\e}{\varepsilon}
\newcommand{\ee}{\varepsilon^{'}}
\newcommand{\s}{{\mit{\it \Sigma}}}
\newcommand{\J}{\mbox{\boldmath$J$}}
\newcommand{\vv}{\mbox{\boldmath$v$}}
\newcommand{\Jh}{J_{{\rm H}}}
\newcommand{\LL}{\mbox{\boldmath$L$}}
\renewcommand{\SS}{\mbox{\boldmath$S$}}
\newcommand{\Tc}{$T_{\rm c}$ }
\newcommand{\Tcf}{$T_{\rm c}$}
\newcommand{\etal}{{\it et al.}: }
\newcommand{\Co}{${\rm Na_{x}Co_{}O_{2}} \cdot y{\rm H}_{2}{\rm O}$ }
\newcommand{\Cof}{${\rm Na_{x}Co_{}O_{2}} \cdot y{\rm H}_{2}{\rm O}$}

 Since the discovery of superconductivity in the non-centrosymmetric 
heavy Fermion compound
CePt$_3$Si,~\cite{rf:bauerDC} superconductivity in materials 
without inversion center has been attracting growing interest.
Many new non-centrosymmetric superconductors with unusual properties 
have been identified among heavy fermion systems
such as UIr,~\cite{rf:akazawa} CeRhSi$_3$,~\cite{rf:kimura} 
CeIrSi$_3$,~\cite{rf:CeIrSi} CeCoGe$_3$~\cite{rf:CeCoGe} and others like
Li$_2$Pd$_{x}$Pt$_{3-x}$B,~\cite{rf:togano} 
and KOs$_2$O$_6$~\cite{rf:shibauchiKOO}.
One immediate consequence of non-centrosymmetricity is
the necessity for a revised classification scheme of 
Cooper pairing states, as parity is not 
available as a distinguishing symmetry. 
The pairing states is considered as mixtures of states with 
different parity 
imposed by the presence of antisymmetric 
spin-orbit coupling (ASOC).~\cite{rf:edelsteinMixChi}  
 Recent theoretical studies let to the proposal of 
various intriguing properties of such a superconductor.~\cite{
rf:edelsteinMixChi,rf:gorkov,rf:frigeri,
rf:fujimoto,rf:kaurHV,rf:samokhin,rf:hayashiSD}

 In the past the relation between superconductivity and magnetism has been 
one of the aspects of major interest in heavy fermion systems. 
Interestingly, all presently known non-centrosymmetric heavy Fermion 
superconductors, i.e. CePt$_3$Si, UIr, CeRhSi$_3$, CeIrSi$_3$ and CeCoGe$_3$,  
coexist with the magnetism.  
 Although magnetism affects 
the electronic state profoundly,  most of the theoretical studies except for 
Refs.~14 and ~15 neglected this aspect so far. 
 The aim of the present study  is to elucidate how the magnetism influences the
superconducting (SC) phase and how it may be involved in deciding the 
pairing symmetry in  CePt$_3$Si. 
Among the non-centrosymmetric heavy fermion superconductors, 
CePt$_3$Si has been investigated in most detail because the superconductivity 
occurs at ambient pressure.~\cite{rf:bauerDC} 
Although we focus here on CePt$_3$Si,  we believe
that some of our results are qualitatively valid for the other compounds too. 

In CePt$_3$Si  superconductivity with $T_{\rm c}=0.75$K appears 
in the antiferromagnetic (AFM) state with Neel temperature 
$T_{\rm N}=2.2$K.~\cite{rf:bauerDC} 
Neutron scattering measurements characterize the AFM order 
with an ordering wave vector $Q=(0,0,\pi)$ and magnetic moments in the 
{\it ab}-plane of the tetragonal crystal lattice.~\cite{rf:metoki} 
The nature of the SC phase has been characterized by several experiments.
The low-temperature properties of the thermal conductivity,~\cite{rf:izawa} 
superfluid density~\cite{rf:bonalde} and specific heat~\cite{rf:takeuchiC} 
indicate line nodes in the gap, while the coherence peak in NMR 
$1/T_{1}T$ is a feature expected rather for a conventional 
superconductor.~\cite{rf:yogiT1} 
The upper critical field $H_{\rm c2} \sim 4$T exceeds  the standard 
paramagnetic 
limit~\cite{rf:bauerDC}, which seems to be consistent with the 
Knight shift data displaying  no decrease of the spin susceptibility 
below $T_{\rm c}$ for any field direction.~\cite{rf:yogiK,rf:higemoto} 
The combination of all these features is incompatible with the usual
pairing states such as the $s$-wave, $p$-wave or $d$-wave state, 
and calls for an extension of the standard working scheme.

For the following study of superconductivity in CePt$_3$Si, 
we introduce the single-orbital Hubbard model including AFM order 
and ASOC, expressed as
\begin{eqnarray}
\label{eq:Hubbard-model}
&& \hspace*{-5mm}  H = \sum_{k,s} \e(\k) c_{\k,s}^{\dag}c_{\k,s} 
   + \alpha  \sum_{k,s,s'} \vec{g}(\k) \cdot \vec{\sigma}_{ss'} 
                            c_{\k,s}^{\dag}c_{\k,s'} 
\nonumber \\
&& \hspace*{-5mm}   - \sum_{k,s,s'} \vec{h}_{\rm Q}  \cdot \vec{\sigma}_{ss'} 
                            c_{\k+\Q,s}^{\dag}c_{\k,s'} 
   + U \sum_{i} n_{i,\uparrow} n_{i,\downarrow}. 
%
\end{eqnarray}
We consider a simple tetragonal lattice and assume the 
dispersion relation as,
\begin{eqnarray}
\label{eq:dispersion}
&& \hspace*{-8mm}  \e(\k)  =   2 t_1 (\cos k_{\rm x} +\cos k_{\rm y}) 
         + 4 t_2 \cos k_{\rm x} \cos k_{\rm y} 
\nonumber \\ && \hspace*{-8mm} 
         + 2 t_3 (\cos 2 k_{\rm x} +\cos 2 k_{\rm y})
+ [ 2 t_4 + 4 t_5 (\cos k_{\rm x} +\cos k_{\rm y}) 
\nonumber \\ && \hspace*{-8mm} 
         + 4 t_6 (\cos 2 k_{\rm x} +\cos 2 k_{\rm y}) ] \cos k_{\rm z}
         + 2 t_7 \cos 2 k_{\rm z} 
         - \mu, 
\end{eqnarray}
which reproduces the so-called $\beta$-band of CePt$_3$Si as obtained from 
band structure calculations without taking AFM order into 
account.~\cite{rf:samokhinband,rf:anisimov,rf:hashimotoDHV} 
 The $\beta$-band has a substantial Ce 4$f$-electron 
character~\cite{rf:anisimov} and the largest density of states (DOS), 
namely 70\% of the total DOS.~\cite{rf:samokhinband} 
 We determine the chemical potential $\mu$ so that the electron density per 
site is $n$ and the parameters as 
$(t_1,t_2,t_3,t_4,t_5,t_6,t_7,n) = (1,-0.15,-0.5,-0.3,-0.1,-0.09,-0.2,1.75)$ 
defining $ t_1 $ as the unit energy.

The second term in eq.~(1) describes the ASOC due to the lack of 
inversion symmetry. 
In case of CePt$_{3}$Si 
the $g$-vector has the Rashba type structure.~\cite{rf:gorkov,rf:frigeri} 
Although the detailed momentum dependence of the $g$-vector 
is not easily obtained from band structure calculations, 
it can reasonably be expressed in terms of velocities 
$v_{\rm x,y}(\k) = \partial \e(\k)/\partial k_{\rm x,y}$ : 
$\vec{g}(\k)= (- v_{\rm y}(\k) , v_{\rm x}(\k), 0) /\bar{v}$. 
 We normalize $\vec{g} $ by the average velocity $\bar{v}$ 
[$\bar{v}^{2}=\frac{1}{N}\sum_{k}v_{\rm x}(\k)^{2}+v_{\rm y}(\k)^{2}$]. 
This form reproduces the correct symmetry and periodicity of 
the $ \vec{g}(\k) $ within the Brillouin zone.
 We choose the coupling constant $\alpha=0.3$ so that a band splitting 
due to ASOC is consistent with the band structure 
calculations.~\cite{rf:samokhinband} 
 Figure~1 shows the Fermi surfaces in our model. 

 The AFM order enters in our model through the staggered field 
$\vec{h}_{\rm Q}$ without discussing its microscopic origin. 
The phase diagram under pressure implies that 
the AFM order is mainly carried by 
localized Ce 4$f$-electrons which have a 
character different from the SC quasiparticles. 
The SC \Tc is little affected by the AFM order which 
vanishes at $P \sim 0.6$GPa~\cite{rf:tateiwa} 
in contrast to the other Ce-based superconductor.~\cite{rf:kitaoka} 
 The experimentally determined order corresponds to $ \vec{h}_{\rm Q} $ 
pointing along the $x$-direction with a wave vector 
$ \Q = (0,0, \pi ) $.~\cite{rf:metoki}  
 For the magnitude we choose $h_{\rm Q} \ll W $ where $W$ is the band width 
since the observed moment $\sim 0.16 \mu_{\rm B}$ is considerably less 
than the full moment of the Ce-ion.~\cite{rf:metoki} 
 We do not touch the complex heavy Fermion aspect, 
i.e. the hybridization of the conduction electrons with the Ce 4$f$-electrons 
forming the strongly renormalized quasiparticles. However we consider the
Hubbard model as a valid effecive model to describe the low-energy 
quasiparticles.~\cite{rf:yanaseReview}

 The undressed Greens function for $U=0$ has the matrix form, 
$
\hat{G}(\k,{\rm i}\omega_{n}) = 
({\rm i}\omega_{n} \hat{1} - \hat{H}(\k))^{-1},
$
where 
\begin{eqnarray}
\label{eq:Green-function}
\hat{G}(\k,{\rm i}\omega_{n}) = 
\left(
\begin{array}{cc}
\hat{G}^{1}(\k,{\rm i}\omega_{n}) & 
\hat{G}^{2}(\k,{\rm i}\omega_{n})\\
\hat{G}^{2}(\kp,{\rm i}\omega_{n}) & 
\hat{G}^{1}(\kp,{\rm i}\omega_{n})\\
\end{array}
\right),  
\end{eqnarray}
%
%
\begin{eqnarray}
\label{eq:H-matrix}
\hat{H}(\k)
= 
\left(
\begin{array}{cc}
\hat{e}(\k) &
-h_{\rm Q} \hat{\sigma}^{({\rm x})} \\
-h_{\rm Q} \hat{\sigma}^{({\rm x})} & 
\hat{e}(\kp) \\
\end{array}
\right), 
\end{eqnarray}
with $\hat{e}(\k)=\e(\k)\hat{\sigma}^{(0)} 
+ \alpha \vec{g}(\k) \vec{\sigma}$ and $\kp = \k+\Q$.  
$\hat{G}^{i}(\k,{\rm i}\omega_{n})$ is a 2 $\times$ 2 matrix in spin space, 
$\omega_{n}=(2 n + 1) \pi T$ 
and $T$ is the temperature.

 We turn to the SC instability which we assume to arise through
electron-electron interaction incorporated in the effective on-site 
repulsion $U$. 
 The linearized \eli equation is obtained in the standard procedure: 
\begin{eqnarray}
\label{eq:Eliashberg}
&& \hspace*{-11mm} \lambda \Delta_{p,s_1,s_2} (\k) = 
- \sum_{\rm k',q,s_3,s_4} V_{p,q,s_1,s_2,s_3,s_4}(\k,\kk) \psi_{q,s_3,s_4}(\kk),
\\
&& \hspace*{-11mm} 
\psi_{p,s_1,s_2}(\k) = \sum_{i,j,s_3,s_4} \phi_{p,i,j,s_1,s_3,s_2,s_4}(\k) 
\Delta_{q,s_3,s_4}(\kkk),
\end{eqnarray}
where $q=p$ ($q=3-p$) for $i=j$ ($i \ne j$), $\kkk=\k+(i-1)\Q$ and 
\begin{eqnarray}
&& \phi_{p,i,j,s_1,s_2,s_3,s_4}(\k) =  
T \sum_{n} G^{i}_{s_1,s_2}(\k,{\rm i}\omega_{n})  
\nonumber \\ && 
\times 
           G^{j}_{s_3,s_4}(-\k + (p-1)\Q,-{\rm i}\omega_{n}) 
\hspace*{5mm} (p=1,2). 
\end{eqnarray}
 Here, we adopt the so-called weak coupling theory of superconductivity 
and ignore self-energy corrections and the frequency dependence 
of effective interaction.~\cite{rf:yanaseReview,rf:miyake}
 This simplification affects the resulting transition temperature but
hardly the symmetry of pairing.~\cite{rf:yanaseReview} 
 The effective interaction 
$V_{p,q,s_1,s_2,s_3,s_4}(\k,\kk)$ originates 
from spin fluctuations which we describe within the RPA.~\cite{rf:miyake}

 The linearized \eli equation allows us to determine the form of 
the leading pairing instability which is attained for the temperature 
at which the largest eigenvalue $ \lambda $ in eq.~(5) reaches unity. 
 We perform the calculation at a given temperature, $ T=0.02 $ 
and determine the most stable pairing state as the eigenfunction of 
the largest eigenvalue for the sake of numerical 
accuracy.~\cite{rf:yanaseReview} 
 The typical value of the eigenvalue at $T=0.02$ and $U=4$ lies around
$\lambda = 0.4 \sim 0.6$.

\begin{figure}[htbp]
  \begin{center}
\includegraphics[width=8.5cm]{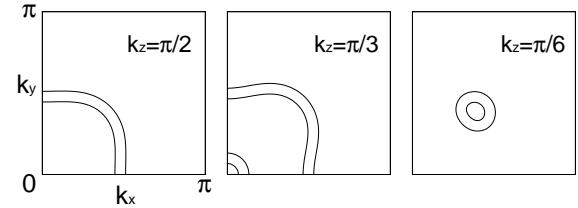}
\caption{
The Fermi surfaces of the Hubbard model (eq.~(1)) at $\alpha=0.3$ 
and $h_{\rm Q}=0$. 
The cross sections at $k_{\rm z}=\frac{\pi}{2}$, 
$k_{\rm z}=\frac{\pi}{3}$ and $k_{\rm z}=\frac{\pi}{6}$ are shown 
from the left to the right. 
}
    \label{fig:fermisurface}
  \end{center}
\end{figure}

First of all, we discuss the symmetry of the SC state which we express by the
following extended parameterization of the gap function:
\begin{eqnarray}
\label{eq:d-vector}
\Delta_{1,s,s'}(\k)
= 
\left(
\begin{array}{cc}
-d_{{\rm x}}(\k)+{\rm i}d_{{\rm y}}(\k) & \Phi(\k) + d_{{\rm z}}(\k) \\
-\Phi(\k) + d_{{\rm z}}(\k) & d_{{\rm x}}(\k)+{\rm i}d_{{\rm y}}(\k) \\
\end{array}
\right), 
\end{eqnarray}
where we use the usual even parity scalar function $ \Phi (\k) $ 
and the odd parity vector $ \vec{d}(\k) $. 
In the presence of  AFM order a second component
$\Delta_{2,s,s'}(\k)$ appears.  
 However, the basic properties and symmetries are little 
affected by $\Delta_{2,s,s'}(\k)$. Within the described scheme 
we identify two stable solutions of the \eli equation 
eqs.~(5-7). One pairing state has dominant $p$-wave symmetry whose order 
parameter has the leading odd parity component 
$\vec{d}(\k) = (-\sin k_{\rm y},\beta \sin k_{\rm x}, 0)$ and the admixed 
even parity part $\Phi(\k) = \cos k_{\rm x}+\cos k_{\rm y}$.
The parameter $\beta$ is unity in the absence of AFM order. 

 The other stable solution has dominantly $d $-wave character
and can be viewed as an inter-layer pairing state: 
$\Phi(\k) = 
\{ \sin k_{\rm x} \sin k_{\rm z} , \sin k_{\rm y} \sin k_{\rm z} \} $ 
(two-fold degenerate) admixed with odd-parity component  
$\vec{d}(\k) =  \Phi(\k) (\sin k_{\rm y} , \sin k_{\rm x},0 ) $.
 In the paramagnetic phase the most stable combination of 
the two degenerate states is chiral: 
$ \Phi_{\pm} (\k) = (\sin k_{\rm x} \pm i \sin k_{\rm y}) \sin k_{\rm z} $ 
which gains the maximal condensation energy 
in the weak-coupling approach. 
In the AFM state, however, the two states of $ \Phi (\k) $ 
are no longer degenerate. 

 A brief view on the pairing mechanism clarifies the origin of 
the stable pairing states. The static spin susceptibility 
is peaked around $ \Q = (0,0, \pi ) $ consistent with  
the AFM order of the Ce moments.
 The in-plane ferromagnetic correlations favor the in-plane $p$-wave pairing. 
 On the other hand, the interlayer AFM correlation drives
interlayer spin singlet pairing. 
 The stability of the two states is determined by the band filling $ n $ 
and the Coulomb repulsion $ U $. Small $U$ and large $ n $ favor the $p$-wave 
state, while the $d$-wave state is more stable for large $ U $ and 
small $n $. The two states are essentially degenerate for $ U =4 $ 
and $ n=1.75 $.

\begin{figure}[htbp]
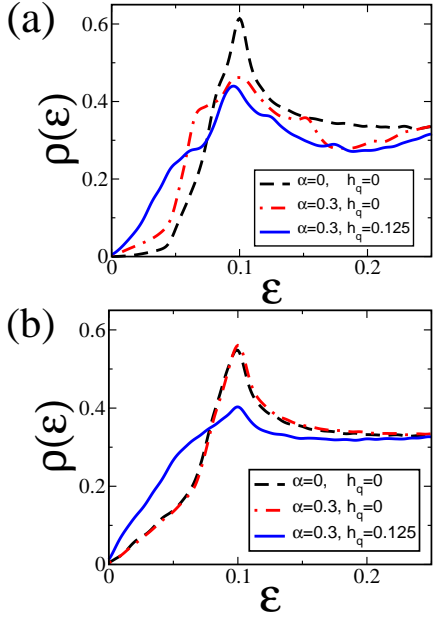

  \begin{center}
\includegraphics[width=5.7cm]{fig2a.eps}
\hspace{5mm}
\includegraphics[width=5.7cm]{fig2b.eps}
\caption{(Color online)
DOS $\rho(\e)$ for $U=4$ in  
(a) dominantly $p$-wave state and (b) dominantly $d$-wave state. 
We show the results for $\e > 0$ because $\rho(\e)$ is particle-hole 
symmetric owing to its definition. 
}
\label{fig:dos}
  \end{center}
\end{figure}

Next we turn to the influence of the AFM order on these pairing states. 
As a first point we discuss the quasiparticle DOS 
which is obtained by diagonalizing the 8 $\times$ 8 matrix, 
\begin{eqnarray}
\label{eq:8-8-matrix}
\hat{H}_{\rm s}(\k)
=
\left(
\begin{array}{cc}
\hat{H}(\k) & -\hat{\Delta}(\k) \\
-\hat{\Delta}^{\dag}(\k)  & -\hat{H}(-\k)^{\rm T} \\
\end{array}
\right), 
\end{eqnarray}
where
\begin{eqnarray}
\label{eq:Delta-matrix}
\hat{\Delta}(\k) 
= 
\left(
\begin{array}{cc}
\Delta_{1,s,s'}(\k) & \Delta_{2,s,s'}(\k) \\
\Delta_{2,s,s'}(\kp) & \Delta_{1,s,s'}(\kp) \\
\end{array}
\right). 
\end{eqnarray}
 The DOS is obtained by the eigenvalues as 
$\rho(\e) = \frac{1}{4N}\sum_{i,k} \delta(\e-E_{i}(\k))$. 
 We determine the momentum and spin dependences of order parameter 
within the linearized \eli equation at $T=0.02$ and assume that those 
structures do not change below \Tcf. 
 For our purpose it is not necessary to calculate the magnitude of the gap 
functions self-consistently as we are mainly interested on qualitative 
properties arising from the gap structure. 
 Thus we choose the magnitude of the maximal gap, $\Delta_{\rm g}=0.1$, 
which may be large compared to the energy scales 
$ \alpha $ or $ h_{\rm Q} $. However, we adopt this value for 
the sake of numerical accuracy, having confirmed that lower values 
of $ \Delta_{\rm g} $ do not alter the result in a qualitative way.

 We first consider the $p$-wave state (Fig.~2(a)). In the absence of ASOC 
and AFM order there are only point nodes along the [001]-direction 
leading to a quadratic energy dependence of the DOS: 
$\rho(\e) = c_1 \e^{2}$.
  The inclusion of ASOC ($ \alpha =0.3 $) yields 
two kinds of the line node. The admixture of the $s$-wave component is one
cause of line nodes as discussed by Frigeri et al.~\cite{rf:frigeri}. 
 The other origin of nodes lies in the specific structure of the $g$-vector.
For the assumed band structure and
$g$-vector, singularities of $\vec{g} $ appear not only
along the [001]-direction (given by symmetry) but also accidentally on lines 
around $(k_{\rm x},k_{\rm y})=(\pm 0.4\pi, \pm 0.4\pi)$. 
 The SC $p$-wave gap vanishes along the lines where 
$\vec{d}(\k) \perp \vec{g}(\k)$.~\cite{rf:yanaseFull} This second type of
line nodes, however, depends strongly on details of material parameters.
Within our model the length of the line nodes 
arising from these mechanism are short leading only to  a weak 
linear energy dependence of the DOS, $\rho(\e) =c_2 \e$ as can be 
seen in Fig.~2(a). 
The linear DOS increases remarkably through the AFM order
(Fig.~2(a)), caused by two effects:  
(I) the Brillouin zone folding at $k_{\rm z}=\pm \pi/2$; 
(II) the modification of SC order parameter. 
Effect (I)  has been investigated by Fujimoto.~\cite{rf:fujimoto} 
It turns out that this effect
is of minor quantitative importance, if $h_{\rm Q} \ll W$ as 
in the present situation.
 Hence the main effect originates from (II), because the  $d_{\rm x}$- and 
$d_{\rm y}$-components are no longer equivalent in the AFM state. 
 The anisotropy parameter 
$ \beta $ in $\vec{d}(\k) = (-\sin k_{\rm y},\beta \sin k_{\rm x}, 0)$ 
decreases with growing $h_{\rm Q}$. In this way the SC gap becomes 
anisotropic leading to the extension of the line nodes and an even
more pronounced linear energy dependence of the DOS. 
 It should be noted that the line nodes discussed for this case
are not symmetry protected but ''accidental''.

 The $d$-wave state has already line nodes for symmetry reasons. 
 In this case the low-energy DOS is not so strongly affected by 
the ASOC as shown in Fig.~2(b). 
 But the slope increases, if the AFM order is included, since the 
pairing state changes its form from 
$d_{\rm xz} \pm {\rm i} d_{\rm yz}$ to $d_{\rm xz}$. 
 The latter has obviously more line nodes.

\begin{figure}[ht]
\begin{center}
\includegraphics[width=6cm]{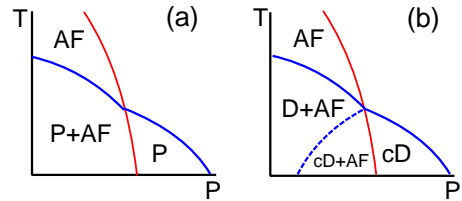}
\caption{(Color online)
Schematic phase diagram in the $P$-$T$ plane. 
(a) $p$-wave and (b) $d$-wave state. 
``D'' (``cD'') shows the $d_{\rm xz}$-wave  
($d_{\rm xz} \pm {\rm i}d_{\rm yz}$-wave) state. 
 The critical temperatures of SC and AFM orders are written 
so as to be consistent with the experiment.~\cite{rf:tateiwa} 
} 
\label{multiplephase}
\end{center}
\end{figure}

 Consequently, the low energy excitations are increased by the AFM order 
for both the $p$-wave and $d$-wave states. 
 The resulting line node behavior is consistent with the experimental 
results at ambient pressure.~\cite{rf:izawa,rf:bonalde,rf:takeuchiC}

The $p$-wave case leads to a simple phase diagram in the $P$-$T$ plane 
(Fig.~3(a)). 
 However, the situation is more intriguing for the $d$-wave case, 
since there is 
an additional phase transition line meeting at the crossing point of 
$ T_N $ and $ T_c $. 
 Generally we would expect an additional SC phase transition within
the SC phase leading  to chiral
$d$-wave phase at low enough temperatures, 
while the SC high-temperature phase 
has only a finite $d_{\rm xz} $-component (Fig.~3(b)). 
Although a second SC transition has been observed,~\cite{rf:nakatsuji} 
it remains unclear whether it represents an intrinsic property or is caused by
sample inhomogeneity.

\begin{figure}[ht]
\begin{center}
\includegraphics[width=5.6cm]{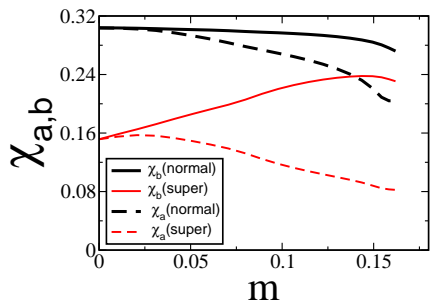}
\caption{(Color online)
Magnetic susceptibility along the $a$- (dashed) and $b$-axis (solid) 
against the staggard spin polarization 
$m=|<\sum_{s,s'}\sigma^{\rm (x)}_{ss'} c_{i,s}^{\dag}c_{i,s'} >|$. 
The thick and thin line show the results in the normal state and SC state 
at $T=0$, respectively. 
} 
\label{susceptibility}
\end{center}
\end{figure}

 It has been reported that the Knight shift remains 
constant below \Tc for any field direction.~\cite{rf:yogiK,rf:higemoto} 
This result looks puzzling in view of calculations 
which suggest the decrease of spin susceptibility for fields 
in the {\it ab}-plane to the half of its normal state 
value.~\cite{rf:edelsteinMixChi,rf:samokhin,rf:gorkov,rf:frigeri} 
 Here, we point out that this discrepancy can be resolved by taking into 
account the AFM order. The uniform spin susceptibility of the normal and 
SC states at $T=0$ is
shown in Fig.~4, assuming $T_{\rm c} \ll \alpha$. 
No correlation effects have been taken into account here. 
For fields $H \perp h_{\rm Q}$ the normal state and SC state susceptibility 
merge for increasing staggered moment.  
 On the other hand, for $H \parallel h_{\rm Q}$ the behavior is opposite. 
Assuming that the anisotropy energy is sufficiently small, the condition 
$H \perp h_{\rm Q}$ is generally favored. Together with other effects 
such as vortex scattering and the formation of a helical SC 
phase~\cite{rf:kaurHV} the influence of AFM order 
could eventually account for the experimental 
results.~\cite{rf:yogiK,rf:higemoto} 
Note that  Fig.~4 is obtained without taking into account 
the canting of AFM moment in contrast to Ref.~15 
where the AFM coupling to the local moment is assumed to explain a similar 
property. Another mechanism to enhance the spin susceptibility below \Tc 
has been proposed by Fujimoto, which is based on a strong particle-hole 
asymmetry.~\cite{rf:fujimoto} However the $ \beta $-band in our model
eq.~(1) does not satisfy the necessary conditions. 
If the AFM order is the main cause for the behavior of 
the spin susceptibility for in-plane fields, a distinct change should be 
observed when the AFM order is suppressed by pressure.

 In summary, we have examined various aspects of the pairing state 
in CePt$_3$Si based on a Hubbard model including the ASOC and AFM order.
For this purpose we chose the $\beta$-band of CePt$_3$Si, 
and found based on spin fluctuation mediated pairing interaction that the 
in-plane $p$-wave and inter-plane $d$-wave states are most likely candidates. 
Both states show  line node behavior, consistent with 
experiments at ambient pressure.~\cite{rf:izawa,rf:bonalde,rf:takeuchiC} 
 The AFM order plays an important role in various respects: 
 (I) the SC gap structure is remarkably deformed 
through the AFM staggered moment.  
 (II) The AFM order can give rise to 
multiple SC phase transitions.
 (III) The in-plane magnetic susceptibility in the SC state can be
increased giving an explanation for the Knight shift 
measurements.~\cite{rf:yogiK,rf:higemoto} 
 The $p$-wave state is more likely realized in CePt$_3$Si because 
it can explain the coherence peak in the NMR 
$1/T_{1}T$.~\cite{rf:yogiT1,rf:hayashiSD,rf:fujimoto} 
 Finally, our results suggest that the investigation of the SC phase under 
the pressure would be interesting, in order to explore the 
SC phase in the regime where AFM order does no longer exist.

 The authors are grateful to D. F. Agterberg, S. Fujimoto, J. Flouquet, 
N. Hayashi, K. Izawa, Y. Matsuda, V. P. Mineev, H. Mukuda, 
T. Shibauchi, R. Settai, H. Tanaka, T. Tateiwa and M. E. Zhitomirsky 
for fruitful discussions.  This study has been 
supported by the Nishina Memorial Foundation, the Swiss Nationalfonds and 
the NCCR MaNEP. 
 Numerical computation was carried out 
at the Yukawa Institute Computer Facility.


\begin{thebibliography}{9}
%

\bibitem{rf:bauerDC}
E. Bauer {\it et al.}: Phys. Rev. Lett {\bf 92} (2004) 027003; 
J. Low. Temp. Phys. {\bf 31} (2005) 748. 

\bibitem{rf:akazawa}
T. Akazawa {\it et al.}: 
J. Phys. Soc. Jpn. {\bf 73} (2004) 3129. 


\bibitem{rf:kimura}
N. Kimura {\it et al.}: 
Phys. Rev. Lett. 95 (2005) 247004.



\bibitem{rf:CeIrSi}
I. Sugitani \etal
J. Phys. Soc. Jpn. {\bf 75} (2006) 043703. 



\bibitem{rf:CeCoGe}
R. Settai: private communication. 


\bibitem{rf:togano}
K. Togano \etal 
Phys. Rev. Lett. {\bf 93} (2004) 247004; 



\bibitem{rf:shibauchiKOO}
T. Shibauchi \etal 
Phys. Rev. B {\bf 74} (2006) 220506. 







%
%
%


\bibitem{rf:edelsteinMixChi}
V. M. Edelstein: Sov. Phys. JETP {\bf 68} (1989) 1244;
Phys. Rev. Lett {\bf 75} (1995) 2004;
Phys. Rev. B {\bf 72} (2005) 172501. 



\bibitem{rf:gorkov}
L. P. Gor'kov and E. I. Rashba: Phys. Rev. Lett {\bf 87} (2001) 037004. 








\bibitem{rf:frigeri}
P. A. Frigeri \etal 
Phys. Rev. Lett {\bf 92} (2004) 097001; 
New. J. Phys. {\bf 6} (2004) 115; 
cond-mat/0505108. 




\bibitem{rf:hayashiSD}
N. Hayashi \etal
Phys. Rev. B {\bf 73} (2006) 024504; 
092508.



\bibitem{rf:kaurHV}
R. P. Kaur \etal 
Phys. Rev. Lett {\bf 94} (2005) 137002. 


\bibitem{rf:samokhin}
K. V. Samokhin: 
Phys. Rev. B {\bf 70} (2004) 104521; 
{\bf 72} (2005) 054514; 
Phys. Rev. Lett {\bf 94} (2005) 027004; 
V. P. Mineev and K. V. Samokhin: cond-mat/0612546.


\bibitem{rf:fujimoto}
S. Fujimoto: Phys. Rev. B {\bf 74} (2005) 024515;
J. Phys. Soc. Jpn. {\bf 75} (2006) 083704; 
cond-mat/0605290. 



















%
%
%

\bibitem{rf:shimahara}
H. Shimahara: 
Phys. Rev. B {\bf 72} (2005) 134518.



\bibitem{rf:metoki}
N. Metoki \etal 
J. Phys. Condens. Matter {\bf 16} (2004) L207. 


\bibitem{rf:izawa}
K. Izawa \etal
Phys. Rev. Lett {\bf 94} (2005) 197002. 


\bibitem{rf:bonalde}
I. Bonalde \etal
Phys. Rev. Lett {\bf 94} (2005) 207002. 


\bibitem{rf:takeuchiC}
T. Takeuchi \etal 
J. Phys. Soc. Jpn. {\bf 76} (2007) 014702. 



\bibitem{rf:yogiT1}
M. Yogi \etal 
Phys. Rev. Lett {\bf 93} (2004) 027003. 


\bibitem{rf:yogiK}
M. Yogi \etal 
J. Phys. Soc. Jpn. {\bf 75} (2006) 013709. 


\bibitem{rf:higemoto}
W. Higemoto {\it et al.}: 
J. Phys. Soc. Jpn. {\bf 75} (2006) 124713. 


%
%
%

\bibitem{rf:samokhinband}
K. V. Samokhin \etal 
Phys. Rev. B {\bf 69} (2004) 094514. 


\bibitem{rf:hashimotoDHV}
S. Hashimoto \etal
J. Phys. Condens. Matter {\bf 16} (2004) L287. 


\bibitem{rf:anisimov}
A. Kozhevnikov and V. Anisimov: private communication. 






\bibitem{rf:tateiwa}
T. Tateiwa \etal 
J. Phys. Soc. Jpn. {\bf 74} (2005) 1903. 




\bibitem{rf:kitaoka}
For a review, Y. Kitaoka {\it et al.} J. Phys. Soc. Jpn. {\bf 74} (2005) 186;
J. Flouquet \etal cond-mat/0505713. 

%
%
%


\bibitem{rf:yanaseReview}
Y. Yanase \etal 
Phys. Rep. {\bf 387} (2004) 1. 





\bibitem{rf:miyake}
K. Miyake {\it et al.}: 
Phys. Rev. B {\bf 34} (1986) 6554;
D. J. Scalapino {\it et al.}: 
Phys. Rev. B {\bf 34} (1986) 8190.


\bibitem{rf:yanaseFull}
Y. Yanase and M. Sigrist: in preparation. 



\bibitem{rf:nakatsuji}
K. Nakatsuji \etal 
J. Phys. Soc. Jpn. {\bf 75} (2006) 084717. 





%
%
%











\end{thebibliography}
\end{document}